\documentclass[twocolumn,aps,prl,showpacs,superscriptaddress,notitlepage]{revtex4-1}
\pdfoutput=1

\usepackage{bbm}

\usepackage[babel]{csquotes}
\usepackage{amsmath}
\usepackage{amsfonts}
\usepackage{braket}
\usepackage{graphicx}
\usepackage{amsmath, amssymb, graphics, setspace}
\usepackage[colorlinks=true,citecolor=blue,urlcolor=blue]{hyperref}

\usepackage{bbold}
\usepackage{braket}
\usepackage{amsthm}

\usepackage{subfigure}
\usepackage{mathrsfs}
\usepackage[usenames, dvipsnames]{xcolor}
\usepackage[a4paper,top=3cm,bottom=3cm,left=1.3cm,right=1.3cm]{geometry}
\usepackage{natbib}

\usepackage{commath} 

\begin{document}

\title{Tunable two-photon quantum interference of structured light}
\author{Vincenzo D'Ambrosio}
\email{vincenzo.dambrosio@icfo.eu}
\affiliation{ICFO - Institut de Ciencies Fotoniques, The Barcelona Institute of Science and Technology, E-08860 Castelldefels, Barcelona, Spain}
\author{Gonzalo Carvacho}
\affiliation{Dipartimento di Fisica, Sapienza Universit\`{a} di Roma, I-00185 Roma, Italy}
\author{Iris Agresti}
\affiliation{Dipartimento di Fisica, Sapienza Universit\`{a} di Roma, I-00185 Roma, Italy}
\author{Lorenzo Marrucci}
\affiliation{Dipartimento di Fisica, Universit\`{a} di Napoli Federico II, Complesso Universitario di Monte S. Angelo, 80126 Napoli, Italy}
\author{Fabio Sciarrino}
\email{fabio.sciarrino@uniroma1.it}
\affiliation{Dipartimento di Fisica, Sapienza Universit\`{a} di Roma, I-00185 Roma, Italy}

\begin{abstract}
Structured photons are nowadays an interesting resource in classical and quantum optics due to the richness of properties they show under propagation, focusing and in their interaction with matter. Vectorial modes of light in particular, a class of modes where the polarization varies across the beam profile, have already been used in several areas ranging from microscopy to quantum information.  One of the key ingredients needed to exploit the full potential of complex light in quantum domain is the control of quantum interference, a crucial resource in fields like quantum communication, sensing and metrology.
Here we report a tunable photon-photon interference between vectorial modes of light. We demonstrate how a properly designed spin-orbit device can be used to control quantum interference between vectorial modes of light by simply adjusting the device parameters and no need of interferometric setups. 
 We believe our result can find applications in fundamental research and quantum technologies based on structured light by providing a new tool to control quantum interference in a compact, efficient and robust way.
\end{abstract}

\maketitle

During the last century quantum mechanics has allowed to describe how even a simple object like a semitransparent surface, can reveal unexpected phenomena like the Hong-Ou-Mandel 
effect \cite{HOM}.
In this case indeed light exhibits a behavior that cannot be predicted by classical theory and can only be explained by considering a purely quantum effect called particle-particle interference \cite{Mandel99}.
Today bosonic coalescence is a key ingredient in photonics; it is exploited in many essential tasks: from the characterization of single photon sources to the measurement of the coherence time of photons and  their degree of distinguishability. Moreover, it lies at the core of the generation of NOON states \cite{NOON}, a key tool in quantum metrology \cite{Flamini2015}, of entanglement swapping and teleportation \cite{teleport}, of quantum fingerprinting \cite{finger} and quantum cloning \cite{17,18,19,20,21}. Finally, quantum interference between two or more photons is nowadays a central resource in quantum computation \cite{BS1,BS2,BS3,BS4,10,11,12,13} and quantum optics in general \cite{Pan2012}.

The fundamental reason for the appearance of photon-photon interference lies in the bosonic nature of light and the impossibility to discriminate between one photon or the other when their modes are totally superimposed. In the original Hong-Ou-Mandel experiment, two photons are sent in the two input ports of a symmetric beam splitter and their indistinguishability is tuned by changing their time delay in order to match all their degrees of freedom. When the time delay is set to zero the two photons are completely indistinguishable and their annihilation/creation operators are identical. As a result, the probability of detecting one photon on each output port vanishes as the final state is a superposition state of two photons exiting from the same port: the so-called photon bunching \cite{HOM}. However this effect is not restricted to beam splitters and can be in general observed every time two (or more) particles are mixed in a way that is impossible to discriminate their contribution to the final state \cite{tritter, agne17,mens17}. Since the time delay is just one possible degree of freedom that can be tuned for quantum interference, other control parameters could be adopted, for instance spectral shape, path superposition, spatial modes and polarization. 

Polarization in particular is usually considered to be uniform across the beam profile, however it is also possible to build vectorial beams where the polarization state varies along the transverse plane according to specific geometries. The most commonly exploited vectorial beams are  the ones corresponding to radial and azimuthal polarization. These states belong to a particular class of vector beams (vector vortex beams, VV) with cylindrical symmetry \cite{Zhan} and show peculiar features under strong focusing \cite{Dorn03} 
and symmetry properties, useful for quantum communication \cite{aligfree,Farias}. Radial and azimuthal polarization are just an example of vector beams since the polarization pattern can exhibit in general a complex structure across the beam profile with interesting applications in a plethora of fields ranging form optical trapping to metrology, nanophotonics, microscopy, quantum information, both in classical and quantum regimes  \cite{Roxw2010,Neug14,Fickl14,gear,roadmap,Maur07,Buse16}. Within quantum information, vector beams have been recently adopted to encode, send and store information \cite{qkdfree,parigi}. Despite the large development of vector vortex beams applications in the quantum domain, a fully controlled quantum interference between two photons in complex modes has never been observed up to now.  Here we fill this gap by reporting an experiment where quantum interference between two photons in two different vectorial modes is observed and controlled by adjusting the parameters of a device that acts as a tunable beam splitter for such modes.

In this article, we observe quantum interference in a spin-orbit coupler device (q-plate)\cite{Marr06} between photons in vector vortex modes. The q-plate here acts as a beam splitter for two different co-propagating vector vortex beams. By tuning the two main parameters of the q-plate we modify the amount of quantum interference between the two structured photons and then control the quantum phase in the output state. 
Our results demonstrate the possibility of tunable quantum interference between vectorial modes of light and hence provide a new tool for quantum experiments with structured light to control two-photon quantum interference in a compact, stable and efficient way.

\section{q-plate action in the vector vortex beams space} 
A convenient way to represent vector vortex beams is through ket states $\ket{\pi,\ell}$ where $\pi$ refers to the polarization state and $\ell$ represents a photon carrying $\ell\hbar$ quanta of orbital angular momentum \cite{vvbeams}.
Let us consider a Hilbert space spanned by the basis states $\{\ket{R,+m},\ket{L,-m}\}$ where $R/L$ stand for circular right/left polarization states respectively. These states correspond to light modes with a uniform circular polarization profile in an OAM eigenstate. By considering balanced superpositions of the basis states we obtain linear polarized light modes with the polarization direction varying across the transverse profile of the beam: 
\begin{align}
\ket{\hat{r}_m}&=\frac{1}{\sqrt{2}}(\ket{R,+m}+\ket{L,-m})\\
\ket{\hat{\theta}_m}&=\frac{1}{\sqrt{2}}(\ket{R,+m}-\ket{L,-m})\\
\ket{\hat{a}_m}&=\frac{1}{\sqrt{2}}(\ket{R,+m}+i\ket{L,-m})\\
\ket{\hat{d}_m}&=\frac{1}{\sqrt{2}}(\ket{R,+m}-i\ket{L,-m})
\end{align}

For each order $m$ the VV beam space can be also represented geometrically via an hybrid Poincar\'{e} sphere \cite{milione,Holl11} where the basis states, corresponding to circular polarizations, coincide with the poles while all the linear polarized states lie on the equator. 
When $m=1$ we have the first order vector vortex beams subspace where $\hat{r}_1$ and $\hat{\theta}_1$ correspond to the radial and azimuthal polarization states respectively.

\begin{figure}
\centering
\includegraphics[width=0.49\textwidth,keepaspectratio=true]{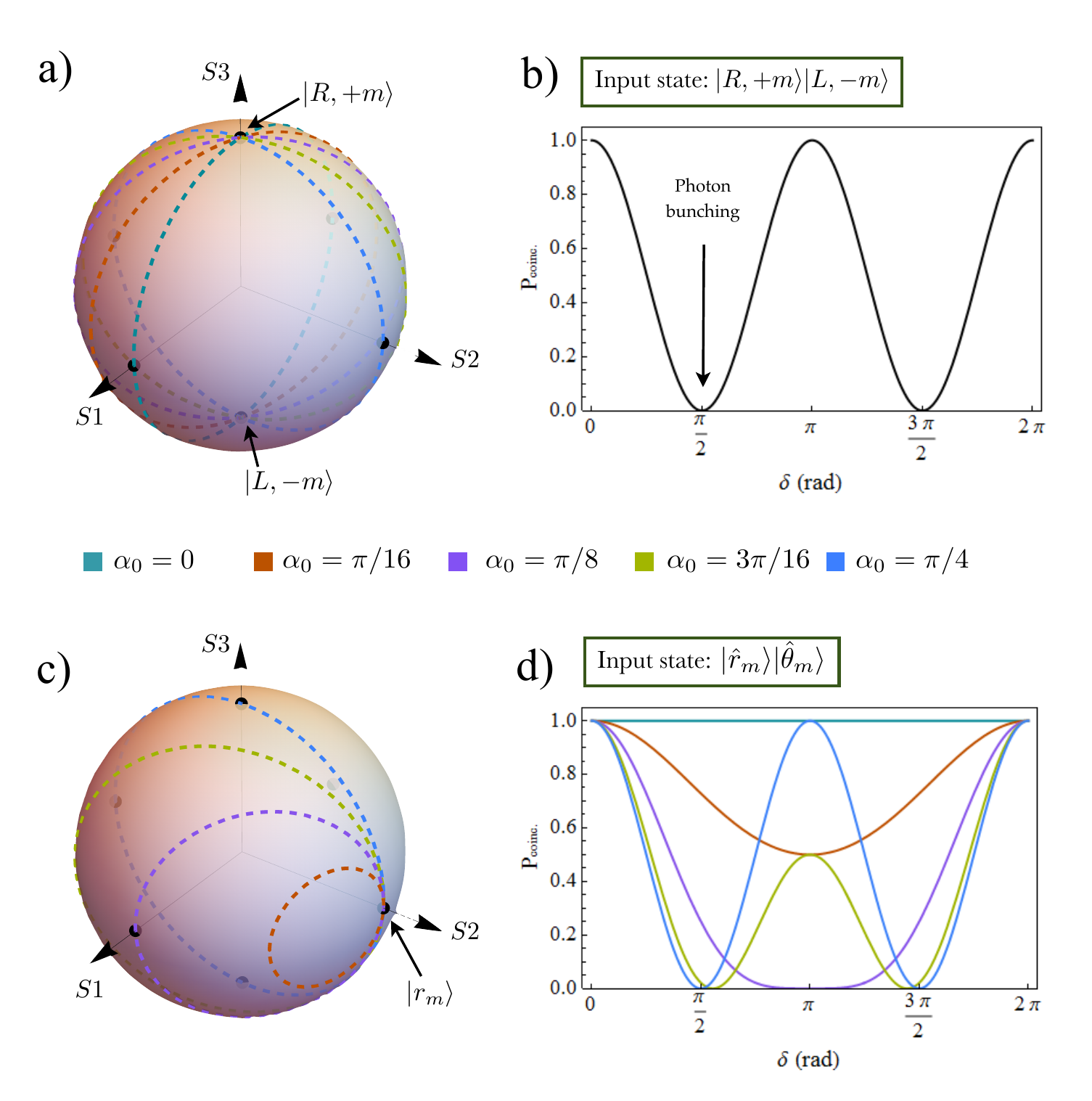}
\caption{\textbf{Two-photon interference in a tunable q-plate:} The transformation induced by a q-plate with $q=m$ on a $m$-th order VV beam can be represented by an orbit on the corresponding hybrid Poincar\'{e} sphere. Different values of $\alpha_0$ correspond to different orientations of the rotation axis in the S1S2 plane while the orbit is explored by changing the retardation $\delta$ of the device. \textbf{a)} When the input photon states lie on the poles the q-plate transformation orbit is a full circle for any $\alpha_0$ value. \textbf{b)} Two photon coincidence probability for the corresponding orbits can be tuned by changing the q-plate phase $\delta$. Such probability does not depend on the offset angle $\alpha_0$. \textbf{c)} Different orbits for different angles $\alpha_0$ for a state lying on the equator (in this case $\ket{r_m}$). Each angle corresponds to a different dependence of the coincidence probability \textbf{d)} respect to the phase $\delta$ introduced by the q-plate.}
 \label{figteo}
\end{figure}

A compact way to generate and measure VV beams is through a q-plate\cite{Marr06}, a liquid crystal inhomogeneous slab where the orientation $\alpha$ of the liquid crystals molecular director depends on the position, represented by the polar coordinates (\textit{r}, $\varphi$), in the following way:
\begin{equation}
\alpha=\alpha_0+ q \varphi
\end{equation}
Here $\varphi$ is the azimuthal angle defined in the slab plane, $q$ is the topological charge of the device and $\alpha_0$ is an offset angle.
The q-plate introduces a phase shift $\delta$ which can be controlled via an externally applied voltage \cite{Piccirillo2010}. 
When $\delta=\pi$ and $q=m/2$ the q-plate acts as an interface between the polarization space and the VV space of order $m$  allowing to generate and measure vector beams by simply acting in the polarization spaces\cite{Cardano2012,vvbeams}.
Interestingly, a q-plate with topological charge $q=m$ allows to operate transformations within the VV space of order m \cite{arbitrary}.
The transformation matrix of this QP in the basis $\{\ket{R,+m},\ket{L,-m}\}$ reads indeed:

\begin{equation}\label{matrix}
QP[\delta,\alpha_0]=
\begin{bmatrix}
\cos\left(\delta/2\right) & i e^{i 2 \alpha_0} \sin\left(\delta/2\right) \\
i e^{-i 2 \alpha_0} \sin(\delta/2) & \cos\left(\delta/2\right)
\end{bmatrix}
\end{equation}
Such transformations can be represented by orbits on the hybrid Poincar\'{e} sphere and can be controlled by tuning the device parameters $\delta$ and $\alpha_0$ (see Fig. \ref{figteo}a,c). 

If we now define $a^\dagger_{1m}(a^\dagger_{2m})$ as the creation operator for a photon in the state $\ket{R,+m}(\ket{L,-m})$, the action of the q-plate reads:

 \begin{equation}
 a^\dagger_{1m} \rightarrow a^\dagger_{1m} \cos\left(\delta/2\right) + a^\dagger_{2m} i e^{-i 2 \alpha_0} \sin\left(\delta/2\right)
 \end{equation} 
 
 \begin{equation}
 a^\dagger_{2m} \rightarrow  a^\dagger_{1m} i e^{i 2 \alpha_0} \sin\left(\delta/2\right) + a^\dagger_{2m}  \cos\left(\delta/2\right) 
  \end{equation}

and the final state is in general a superposition of the input modes.
When $\alpha_0 =0$, the transformation (\ref{matrix}) is formally equivalent to a beam splitter matrix with a tunable reflectivity that depends on $\delta$. In particular, for $\delta=\pi/2$ the q-plate behaves as a symmetric beam splitter. The main difference with a standard beam splitter is that the q-plate does not act on the direction of propagation of light but on two co-propagating structured photons in different vectorial modes. Let us now study the transformation operated by such a q-plate on a two photons input state and check for quantum interference effects:

\begin{equation}\label{scalar}
\begin{split}
 a^\dagger_{1m}a^\dagger_{2m} \ket{00} \rightarrow  a^\dagger_{1m}a^\dagger_{2m} \cos(\delta) \ket{00} +\\+\frac{i}{2} (e^{2 i\alpha_0} (a^\dagger_{1m})^2+ e^{-i 2 \alpha_0} (a^\dagger_{2m})^2 )  \sin(\delta) \ket{00}
\end{split}
\end{equation} 

For $\delta=\pi/2$ (and any value of $\alpha_0$) we observe total photon bunching and the number of coincidences on the two different modes is null (Fig.\ref{figteo}b). Moreover the final state is a NOON state with $N=2$ with a phase that depends on $\alpha_0$ and can then be tuned by simply rotating the q-plate \cite{arbitrary} (the only exception is the case q=1 where the angle $\alpha_0$ needs to be set during the manufacturing process). More in general, the control of the output state via $\delta$ and $\alpha_0$ can be a resource for further quantum information processing when the device is used for instance in a more complex setup like a quantum network.

 To fully characterize the action of the device, let us now consider a photon in a varying linear polarization mode. Without losing generality, we can consider the action of the q-plate on the creation operators $(a^\dagger_{r_m}, a^\dagger_{\theta_m})$ corresponding to the states $\ket{\hat{r}_m}$ and $\ket{\hat{\theta}_m}$
The matrix (\ref{matrix}) in this new basis reads:

\begin{equation}\label{matrix2}
\begin{bmatrix}
\cos\left(\frac{\delta}{2}\right)+i \cos(2\alpha_0) \sin\left(\frac{\delta}{2}\right)& \sin(2 \alpha_0) \sin\left(\frac{\delta}{2}\right)\\
-\sin(2 \alpha_0) \sin\left(\frac{\delta}{2}\right)& \cos\left(\frac{\delta}{2}\right) -i \cos(2\alpha_0) \sin\left(\frac{\delta}{2}\right)
\end{bmatrix}
\end{equation}

A two photon state $a^\dagger_{r_m} a^\dagger_{\theta_m}\ket{0}$ is transformed as follows:
\begin{equation}\label{radiali}
\begin{split}
a^\dagger_{r_m} a^\dagger_{\theta_m} \rightarrow \\ -\sin(2\alpha_0)\sin\left(\frac{\delta}{2}\right)(\cos\left(\frac{\delta}{2}\right)+i \cos(2\alpha_0)\sin\left(\frac{\delta}{2}\right)) (a^\dagger_{r_m})^2 + \\
+\sin(2\alpha_0)\sin\left(\frac{\delta}{2}\right)(\cos\left(\frac{\delta}{2}\right)-i \cos(2\alpha_0)\sin\left(\frac{\delta}{2}\right)) (a^\dagger_{\theta_m})^2+\\
+(\cos^2(2\alpha_0)+\cos(\delta)\sin^2(2\alpha0))a^\dagger_{r_m} a^\dagger_{\theta_m}
\end{split}
\end{equation}
For $\alpha_0=0$ the first two terms vanish and there is no quantum interference, since $a^\dagger_{r_m}$ and $a^\dagger_{\theta_m}$  are eigenstates of the q-plate transformation for $\alpha_0=0$.  Hence these two states are never mixed by varying $\delta$ and they are always completely distinguishable. 
 However, when $\alpha_0\neq0$ the coincidence probability depends on the phase ($\delta$) introduced by the q-plate. Quantum interference can be controlled by tuning q-plate parameters as shown in Fig. \ref{figteo} d where the probability of measuring photon coincidences due to the $a^\dagger_{r_m} a^\dagger_{\theta_m}$ contribution are plotted as a function of the phase $\delta$ of the q-plate for different values of $\alpha_0$. In the particular case of $\alpha_0=\pi/4$ such probability is $P_{r\theta}=cos^2(\delta)$ exactly as for the case of circular polarizations. When total photon bunching occurs, the two initially independent photons, that were in two orthogonal complex polarization modes, are forced to exit the device in the same vector vortex mode.
 
Although the calculation has been carried out for the states $\ket{\hat{r}_m}$ and $\ket{\hat{\theta}_m}$ the q-plate effect can be easily derived for any linear combination of $a^\dagger_{1m}$ and $a^\dagger_{2m}$. As a further example we can consider the states $\ket{\hat{a}_m}$ and $\ket{\hat{d}_m}$ (whose creator operators are $a^\dagger_{a_m}$ and $ a^\dagger_{d_m}$ respectively) that can be obtained by locally rotating the polarization direction by $\pi/4$ in $\ket{\hat{r}_m}$ and $\ket{\hat{\theta}_m}$ respectively and, together with the other two bases here considered, form a complete set of mutually unbiased bases useful in quantum cryptography applications. Due to the geometry of these states, the only difference with respect to the previous case is that the effect of $\alpha_0$ orientation is shifted by $\pi/4$ on the $\alpha_0$ axis.  
\begin{figure*}[htb]
\centering
\includegraphics[width=1\textwidth]{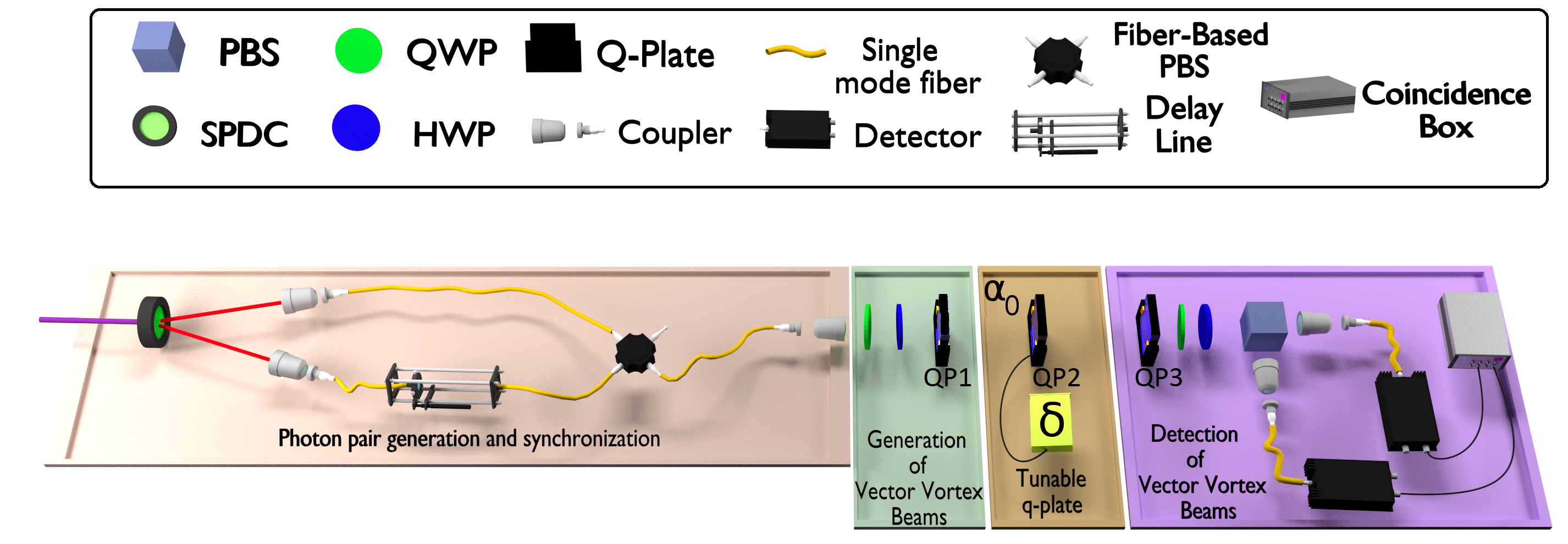}
\caption{\textbf{Experimental apparatus:} In the synchronization stage, photon pairs with orthogonal polarization are generated via Spontaneous Parametric Down Conversion (SPDC) process, synchronized within their coherence time via a delay line and finally coupled into a fiber-based polarizing beam splitter. At the exit port of the fiber-PBS, the two photons are spatially and temporally matched and with orthogonal polarization states.  In the generation stage the photons are then converted into vector vortex state by employing the q-plate "QP1" with topological charge $q=1/2$. Different polarization states (controlled with birefringent waveplates) correspond to the generation of different states in the vector vortex modes space. 
Once vector vortex modes are generated they pass through a second q-plate "QP2" with topological charge $q=1$ where tunable two-photon interference between vectorial modes takes place. This effect is tuned by changing the parameter $\delta$ of the q-plate via an externally controlled voltage, and the offset angle $\alpha_0$ (in this experiment $\alpha_0=0$ and $\alpha_0=\pi/4$). In the detection stage another q-plate "QP3" with topological charge $q=1/2$ converts back the vector vortex modes into uniform polarization states, which are measured with birefringent waveplates and a bulk polarizing beam splitter (PBS). The photons are then coupled into single-mode fibers  
and collected by single-photon detectors connected to a coincidence box.} \label{setup}
\end{figure*} 

\begin{figure*}[ht]
\centering
\includegraphics[width=\textwidth]{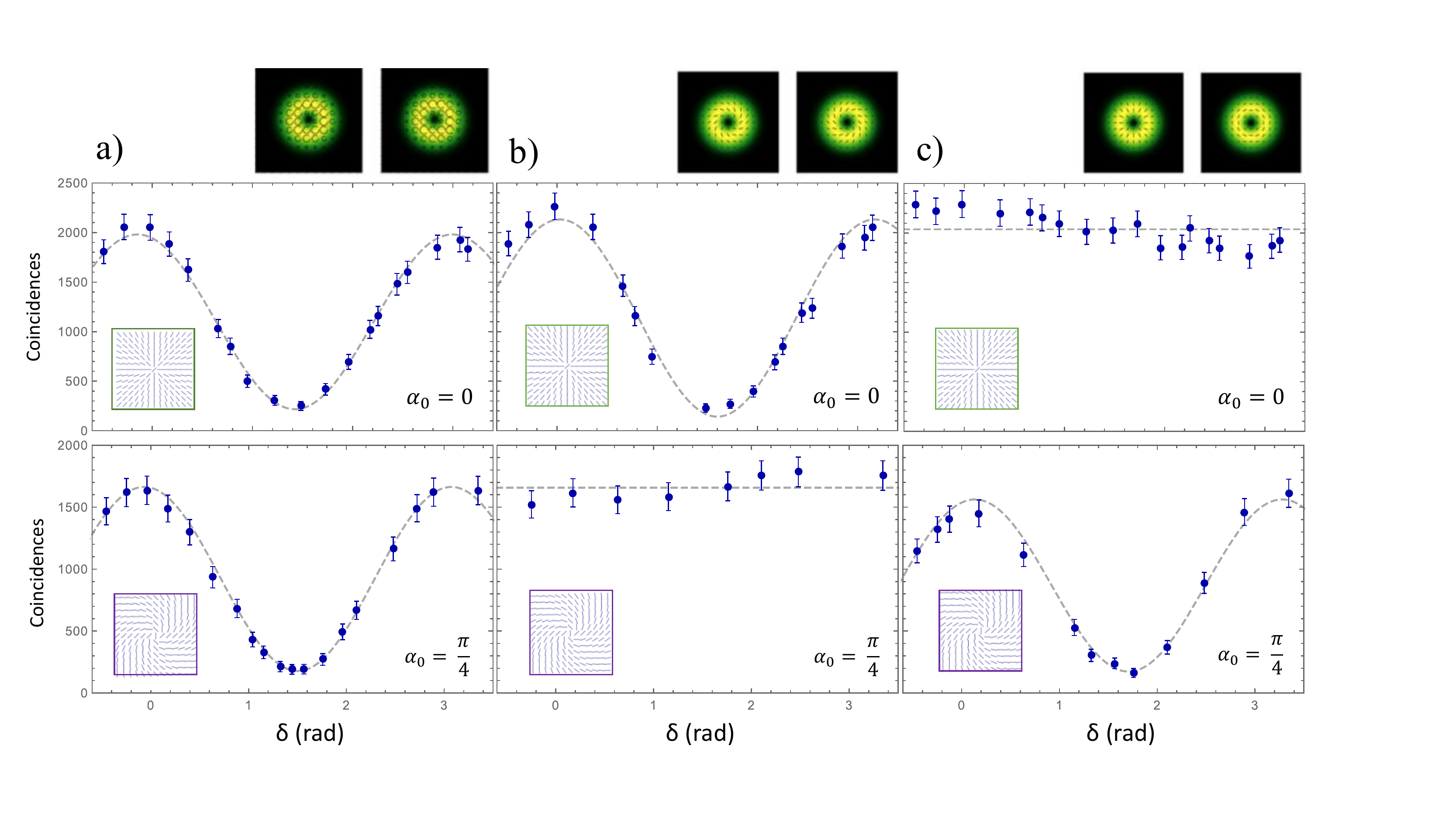}
\caption{\textbf{Experimental tuning of photon bunching through the action of a q-plate:} The six plots show the number of recorded coincidences after two-photon quantum interference takes place in the q=1 q-plate (QP2), as a function of the phase shift $\delta$ introduced by the device. 
The intensity and polarization distribution in the transverse plane for the modes of the two photons is represented in two black boxes in the upper right corner of each column. \textbf{a)}  input state $a_1^\dagger$$a_2^\dagger$$\ket{00}$, with $\alpha_0$=0 (top row, green box) and $\alpha_0$=$\frac{\pi}{4}$ (bottom row, purple box); \textbf{b)} input state $a_a^\dagger$$a_d^\dagger$$\ket{00}$, with $\alpha_0$=0 (top row, green box) and $\alpha_0$=$\frac{\pi}{4}$ (bottom row, purple box); \textbf{c)} input state $a_r^\dagger$$a_\theta^\dagger$$\ket{00}$, with $\alpha_0$=0 (top row, green box) and $\alpha_0$=$\frac{\pi}{4}$ (bottom row, purple box).}
\label{experimental_data}
\end{figure*}

\begin{figure*}[ht]
\includegraphics[scale=0.7]{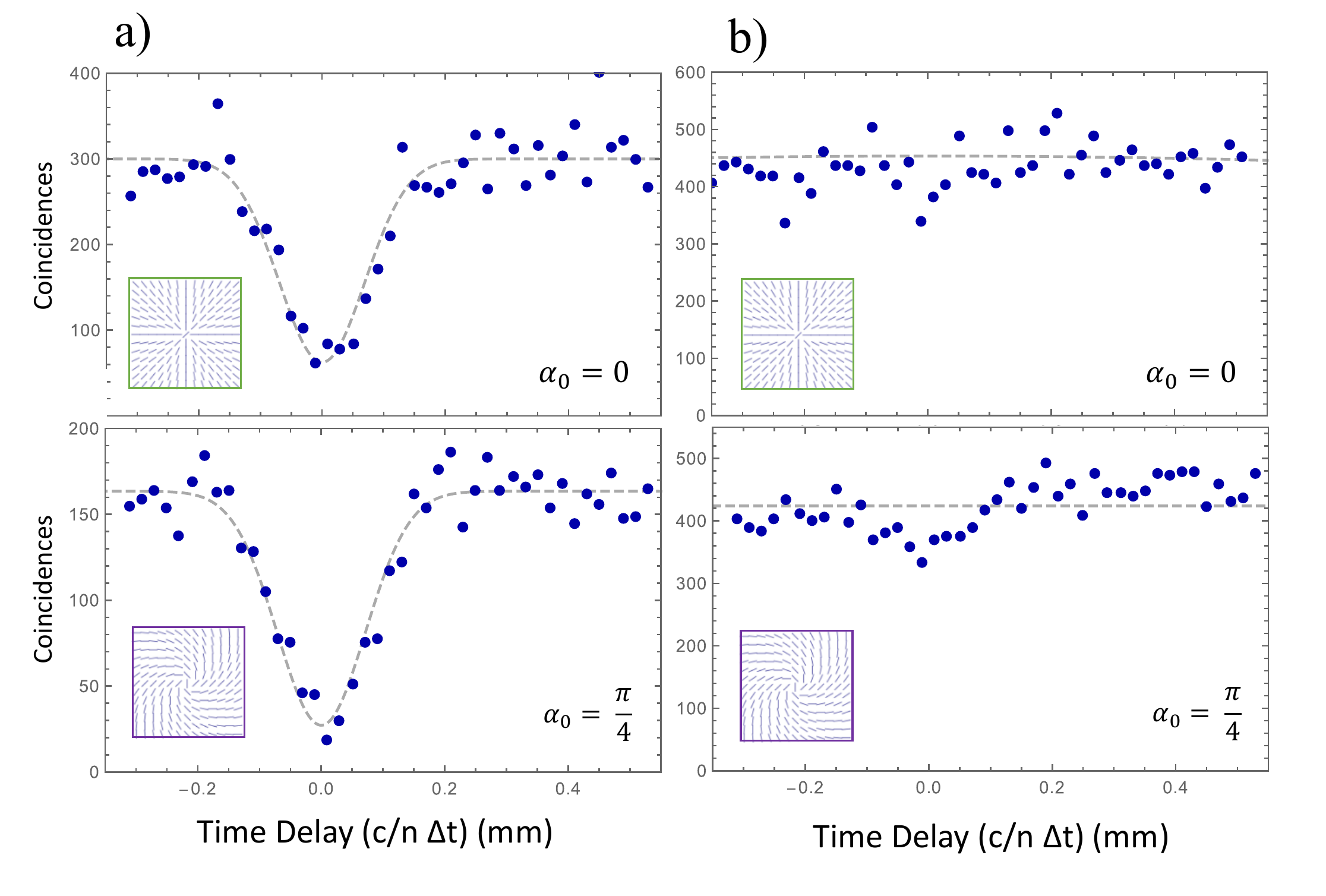}
\caption{\textbf{Delay dependent photon bunching for fixed q-plate tuning $\delta$:} The four 
plots show the number of coincidences in the basis $\{\ket{R+1},\ket{L,-1}\}$ as a function of the time delay between two photons with input state $a_1^\dagger$$a_2^\dagger$$\ket{00}$ 
after they pass through q-plate QP2 
with a fixed phase shift $\delta$. \textbf{a)} $\delta$ is set to induce full quantum interference and coincidences are registered adopting a q-plate with $\alpha_0$=0 (top, green box) and $\alpha_0$=$\frac{\pi}{4}$ (bottom, purple box) 
\textbf{b)} $\delta$ is set to induce no quantum interference and coincidences are registered adopting a q-plate with $\alpha_0$=0 (top, green box) and $\alpha_0$=$\frac{\pi}{4}$ (bottom, purple box).}
\label{HOM_delay}
\end{figure*}

\section{Observation of vector vortex beam bosonic coalescence}
To obtain quantum interference with structured light we generate a photon pair in first order vector vortex beams and control particle-particle interference by tuning the parameters of a q-plate.
More in detail, a photon pair is produced by exploiting spontaneous parametric down conversion process (SPDC). The generated photons are sent through the experimental apparatus depicted in Fig. \ref{setup}. Firstly the two photons are synchronized through a delay line that changes the optical path for one of them in order to control their temporal mismatch $\Delta t$ (see Methods).
Once the photons are temporally synchronized, their polarization 
is rotated, so that both photons end up in the same output of the in-fiber PBS in the state $\ket{H,0}\ket{V,0}$. In the generation stage the vector modes are created by first controlling the two photon polarization state via a quarter wave plate and a half wave plate and then by sending them in a q-plate with topological charge $q=\frac{1}{2}$ (QP1) \cite{vvbeams}. Accordingly, vector vortex modes of order $m=1$ are generated. For the sake of notation simplicity we will omit the subscript $m$ from creation operators  and ket states when referring to first order modes $m=1$. We generated three different inputs: $a^\dagger_1a^\dagger_2\ket{00}$, $a_r^\dagger a^\dagger_\theta\ket{00}$ and $a^\dagger_a a^\dagger_d\ket{00}$.
After the generation, the vector vortex modes are sent through a second q-plate (QP2) with topological charge $q=1$ to perform the transformations described by equations (\ref{matrix}) and (\ref{matrix2}). Here quantum interference takes place and can be tuned by acting on the two parameters $\delta$ and $\alpha_0$ of the device. Since the $q=1$ geometry is the only one for which $\alpha_0$ cannot be tuned by simply rotating the device, we performed the experiment with two different q-plates  one with $\alpha_{0}=\pi /4$ and one with $\alpha_{0}$=0 respectively.
These two q-plates had been previously characterized in order to completely map the external applied voltage into the phase shift $\delta$ \cite{Piccirillo2010}. 
In the final stage of the experiment, the two photons are sent through the detection stage, where a third q-plate with topological charge $q=\frac{1}{2}$ (QP3) together with a polarization analysis setup allows measuring photons coincidences (with two single photon detectors) in an arbitrary basis of the VV beams space \cite{vvbeams}.

The experimental results are reported in Fig. \ref{experimental_data} and Fig. \ref{HOM_delay}. 
The first reports the number of coincidences as a
function of the phase shift $\delta$ introduced by QP2 when the  photons are temporally synchronized $\Delta t=0$. The three columns \textbf{a}, \textbf{b} and \textbf{c} correspond to three different two-photon input states and each row to a different value of $\alpha_0$ namely $\alpha_0=0$ and $\alpha_0=\pi/4$. The geometry of the two q-plates are reported as insets in the plots,  $\alpha_0$=0 (top, green box) and $\alpha_0$=$\frac{\pi}{4}$ (bottom, purple box).
When the input state is $a_1^\dagger a_2^\dagger\ket{00}$ (Fig.\ref{experimental_data}a), quantum interference depends only on the q-plate parameter $\delta$ as predicted in Fig. \ref{figteo}b, showing 
no dependence on $\alpha_0$. The coincidence counts in the basis $\{\ket{R,+1},\ket{L,-1}\}$ are reported in Fig.\ref{experimental_data}a together with the best fit curve according to equation (\ref{scalar}).
When the input photons are in radial and azimuthal states $a_r^\dagger a_\theta^\dagger \ket{00}$  (Fig.\ref{experimental_data}b) the coincidence probability in the $\{\ket{\hat{r}},\ket{\varphi}\}$ basis is constant for $\alpha_0=0$ but 
quantum interference occurs for $\alpha_0=\pi/4$ as predicted by the equation (\ref{radiali}) on which the best fit curves are based.
Finally in Fig.\ref{experimental_data}c 
the coincidence probabilities for $a_a^\dagger a_d^\dagger \ket{00}$ measured in the $\{\ket{a},\ket{d}\}$ basis are shown. As predicted by theory 
the behavior is similar to the radial and azimuthal case but the $\alpha_0$ dependence is shifted by $\pi/4$.

Later, we measured photon coincidences as a function of the temporal delay $\Delta t$ between the two photons 
for fixed values of $\delta$.  The input state was $a_1^\dagger$$a_2^\dagger$$\ket{00}$ and the coincidences were measured in the basis $\{\ket{R,+1},\ket{L,-1}\}$. In Fig. \ref{HOM_delay} are reported the results when $\delta$ is set to induce full quantum interference (column a) and to induce no quantum interference (column b).
The two rows correspond to the two different q-plates used in the experiment: $\alpha_0$=0 (top, green box) and $\alpha_0$=$\frac{\pi}{4}$ (bottom, purple box). We observe the typical Hong-Ou-Mandel dip (photon bunching at $\Delta t =0$) only when quantum interference is enabled by the q-plate action.
All the error bars are obtained by considering a Poissonian photon statistic.

\section{Conclusions and discussions}
Structured photons are becoming an interesting resource for quantum optics due to the richness of properties they show under propagation and in their interaction with matter \cite{roadmap}. However, one of the ingredients needed to exploit the full potential of complex light in quantum domain is the control of quantum interference. 
Here we have reported a tunable photon-photon interference between structured modes of light. Our experiment shows how a properly designed q-plate can be used to control quantum interference between vectorial modes of light by simply adjusting the device parameters. 
This feature can be used for instance in quantum optics experiments with structured light to control the degree of distinguishability between co-propagating photons. One immediate application can be the controlled generation of NOON states of vector beams for quantum metrology applications \cite{gear}. 
In quantum information our result allows performing operations based on quantum interference (entanglement gates, Bell state analysis)  when information is carried by photons in vector modes and with no need of interferometric setups. 
More in general we believe this result provides a new tool for the wide and stimulating area of fundamental research and quantum technologies based on structured light.

\section{Methods}
\textbf{Photon's synchronization:} Photon pairs were generated by a parametric down conversion source, made of a 2 mm thick nonlinear crystal of beta barium borate (BBO) illuminated by a pulsed pump field with $\lambda= 392.5$ nm. After spectral filtering and walk-off compensation, photons are simultaneously sent to an in-fiber polarizing beam splitter (PBS1). The temporal synchronization is achieved exploiting a delay line which is placed on one of the two photons' paths, right before one input of the in-fiber PBS1.\\
To send the input states into the measurement apparatus, the photons' polarizations are compensated in order to make them exit the PBS1 from the same output mode.\\

\section{Acknowledgments}
This work was supported by H2020 European Research Council (ERC) (694683 (PHOSPhOR)).
V.D. acknowledges financial support from the Spanish Ministry of Economy and Competitiveness, through the "Severo Ochoa" Programme for Centres of Excellence in R\&D (SEV-2015-0522)", Fundaci\'{o} Privada Cellex, CERCA Programme / Generalitat de Catalunya and ICFONEST fellowship program. G.C. thanks Becas Chile and Conicyt for a doctoral fellowship.

\end{document}